\documentclass[sigconf,nonacm,10pt]{acmart}

\usepackage{alltt}
\usepackage{multirow}

\definecolor{read}{HTML}{F5793A}
\definecolor{write}{HTML}{0F2080}
\definecolor{execute}{HTML}{A95AA1}

\setcopyright{acmcopyright}
\copyrightyear{2020}
\acmYear{2020}
\acmDOI{10.1145/1122445.1122456}

\acmConference[Under Preparation, do not distribute]{Under Preparation, do not distribute}{June 2020}{Page~\thepage}
\acmBooktitle{Under Preparation, do not distribute}
\acmPrice{15.00}
\acmISBN{978-1-4503-XXXX-X/18/06}

\hypersetup{draft}

\begin{document}

\title[Low-Code Programming Models\hfill\arabic{page}]{Low-Code Programming Models}

\author[Martin Hirzel\hfill\arabic{page}]{Martin Hirzel}
\affiliation{
  \institution{IBM Research}
  \country{USA}}

\begin{abstract}
Traditionally, computer programming has been the prerogative of
professional developers using textual programming languages such as
C, Java, or Python.
Low-code programming promises an alternative: letting citizen developers
create programs using visual abstractions, demonstrations, or natural
language.
While low-code programming is currently getting a lot of attention in
industry, the relevant research literature is scattered, and in fact,
rarely uses the term ``low-code''.
This article brings together low-code literature from various research
fields, explaining how techniques work while providing a unified point
of view.
Low-code has the potential to empower more people to automate tasks
by creating computer programs, making them more productive and less
dependent on scarce professional software developers.
\end{abstract}

\maketitle

\section{Introduction}\label{sec:introduction}

Low-code is the subject of much current enthusiasm stirred by market
research companies and confirmed by vendors rushing to embrace the
label~\cite{bock_frank_2021,sahay_et_al_2020}.
But it is cryptic exactly what low-code programming means, let alone
how it works, and the scientific literature hardly uses the term.
We can decode the term by breaking it into its components.
\emph{Programming} means developing computer programs, which comprise
instructions for a computer to execute.
Traditionally, programming means writing code in a textual programming
language, such as C, Java, or Python.
In contrast, \emph{low-code programming} minimizes the use of a
textual programming language.
Instead, it aims to use alternative techniques that are closer to how
users naturally think about their task.

Users of low-code range from
professional developers to so-called citizen developers.
A \emph{citizen developer} is an amateur programmer with little
professional programming education.
Citizen developers, having chosen a career different from programming,
tend to have more domain expertise.
Low-code enables domain experts to become citizen developers.
At the same time, low-code platforms should also strive to make
\emph{pro-developers}~(professionals with an education or career in software
development) more productive.

Whether used by a citizen developer or a pro-developer, low-code
programming aims to save the time and tedium of performing a
task by hand~\cite{vanderaalst_bichler_heinzl_2018}.
Further motivation for individuals comes from the joy of creating
something useful, thinking about tasks in a computational way, and
acquiring programming skills that can advance their career.
Besides individuals, businesses may have their own motivation for
adopting low-code platforms.
Low-code platforms can alleviate the shortage of pro-developers,
reduce mistakes of tedious manual tasks, and multiply the time savings
from one individual's low-code program to their
colleagues~\cite{sahay_et_al_2020}.
Another factor driving low-code is the rise of cloud-based software as
a service, providing both more interfaces to automate and a platform
on which to deploy automations.

A few concepts are closely related to low-code programming.
\emph{No-code programming} is more purist, with zero hand-written code
in a textual programming language.
This term mostly appears in marketing materials and analyst reports.
\emph{End-user programming}~(EUP) puts the emphasis on who is doing the
programming~(the end-user as citizen developer) rather than on how
they are \emph{not} doing their programming~(not with textual
code)~\cite{barricelli_et_al_2019}.
This term is common in the academic literature and overlaps
with low-code, but does not preclude the use of a textual
programming language.
Another gap between EUP and low-code is that the latter aims to serve
not just end users but also
pro-developers~\cite{bock_frank_2021,sahay_et_al_2020}.

Bock and Frank~\cite{bock_frank_2021} and Sahay et
al.~\cite{sahay_et_al_2020} recently compared commercial low-code
platforms, and Barricelli et al.\ recently
mapped the end-user programming literature~\cite{barricelli_et_al_2019}.
In contrast, this article bridges the gap between low-code and the
academic literature and adds missing details and
perspective.
Low-code encompasses more specialized techniques, such as
VPLs~(visual programming languages), PBD~(programming by demonstration),
PBE~(programming by example), RPA~(robotic process automation),
PBNL~(programming by natural language), and others.
Surveys on these are more specific and often
dated~\cite{androutsopoulos_ritchie_thanisch_1995,boshernitsan_downes_2004,kuhn_2014,vanderaalst_bichler_heinzl_2018}.
In contrast, this article reviews recent literature across all
above-listed techniques.

Given that low-code offers citizen developers a model to create
computer programs, this article explores low-code from the perspective
of programming models.
A \emph{programming model} is a set of abstractions that supports
developing computer programs.
Programming models can be low-code or not, and they can be
domain-specific or general-purpose.
Some programming models are languages; e.g., Java is a general-purpose
language and SQL is domain-specific, and neither is low-code.
Scratch is a low-code programming model for kids that is
media-centric~\cite{resnick_et_al_2009}, making it domain-specific.
The programming-model perspective helps this article highlight common
techniques for writing, reading, and executing programs.
Furthermore, the programming-model perspective helps relate low-code
to research into program synthesis and domain-specific languages.

This article includes a deep-dive for three prominent low-code
techniques: visual programming, programming by de\-monstration, and
programming by natural language.
The deep-dive focuses on fundamental building blocks and a unifying
framework common to all three.
The citations in this article cover both seminal work and recent
advances in low-code programming models, for instance, based on
artificial intelligence.
Overall, this article aims to cut through the buzz surrounding low-code
so as to expose the technical foundations underneath.
We hope that doing so will foster better development of the field through
awareness of existing~(albeit scattered) research.
Ultimately, we hope this will lead to even more empowered citizen
developers.

\section{Problem Statement}\label{sec:problem}

If low-code is the solution, then what is the problem?
Given the term low-code, it might seem that the answer is obviously
code.
Unfortunately, that answer is superficial and non-constructive.
Defining a thing solely by what it is not, as the term low-code
appears to do, causes confusion.
Consider two other recent similarly-named trends:
NoSQL and serverless.
At the surface, one might think NoSQL was mostly about rejecting SQL,
but in fact, it was more about flexible data and consistency models
than about the query language.
Similarly, serverless computing was not about eliminating compute
servers, but about hiding them behind better abstractions.
Defining a new trend by rejecting an old one grabs attention at the
expense of being misleading.
Just like serverless still needs servers, low-code~(and even no-code!)
still needs code.

Instead, the three terms low-code, NoSQL, and serverless have one
thing in common: a desire to avoid specific baggage while
preserving core value.
In NoSQL, the core value is durable and consistent storage.
In serverless, it is portable and elastic compute.
What then is the core value that low-code aims to preserve?
This article argues that it is computer programming.
Programming is to low-code what computing is to serverless.
Low-code is about creating instructions for a computer
to execute or interpret.
These instructions form a computer program, typically in a
domain-specific language~(DSL).
For instance, low-code is often based on search-based program
synthesis, and synthesis usually targets a DSL carefully crafted for
the purpose~\cite{alur_et_al_2018}.
The program may not be exposed to the user, but it is there.

\begin{figure}
\centerline{\includegraphics[width=\columnwidth]{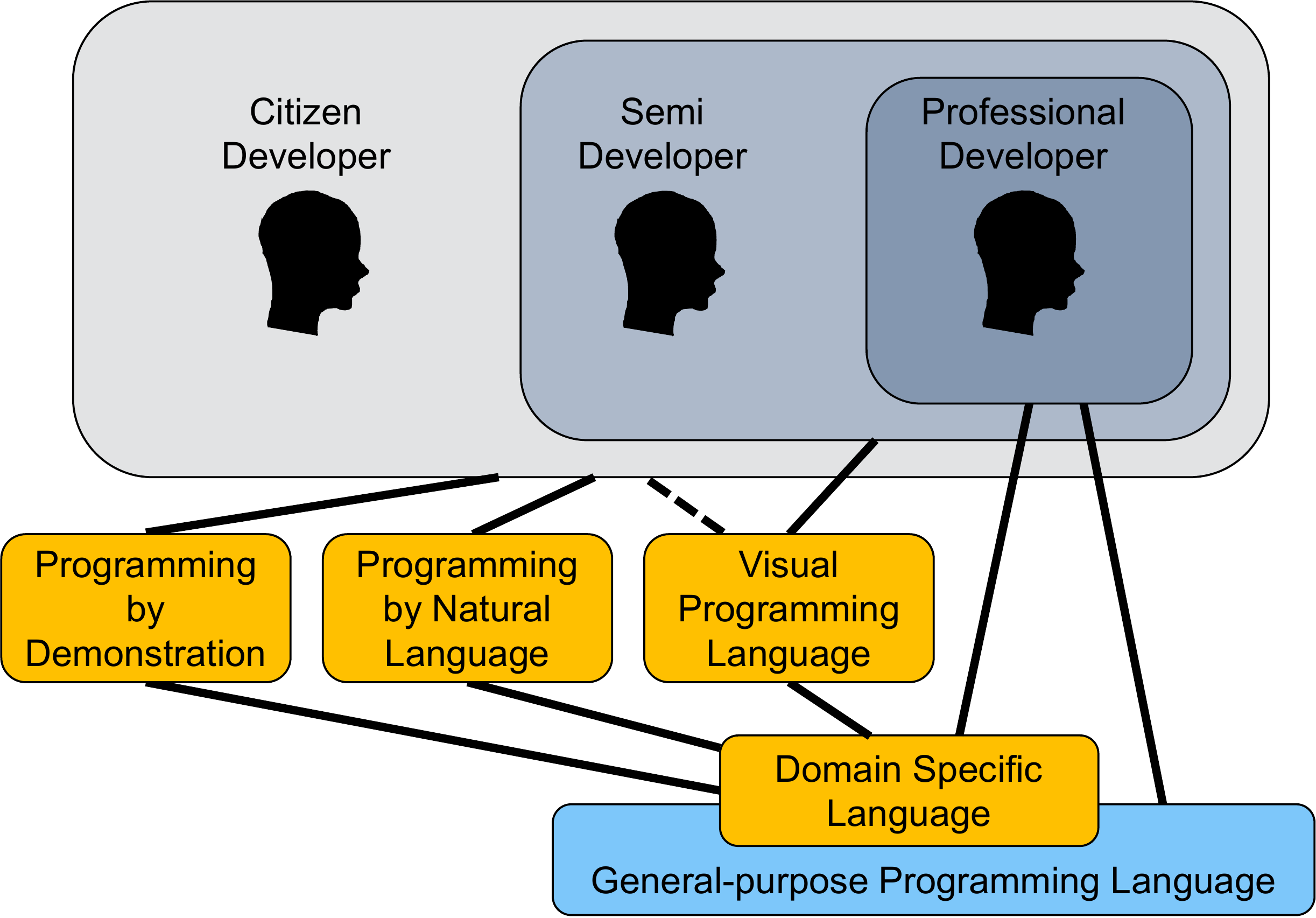}}
\caption{\label{fig:personas}Low-code users and techniques.}
\end{figure}

One way to better understand the problem statement behind low-code is
to look at who it is for.
The top portion of Figure~\ref{fig:personas} shows the spectrum
of low-code users.
They range from citizen developers at one end to pro-developers at the
other end, with intermediate stages here dubbed \emph{semi developers}.
In this simplified view, users at the citizen developer end of the
spectrum tend to have the most domain knowledge and users at the
pro-developer end have the most programming expertise.
Low-code can enable citizen developers to self-serve their programming
needs instead of depending on pro-developers.
At the same time, low-code can make pro-developers more productive,
e.g., in a new domain.
Finally, low-code can break barriers between developers across the
spectrum and help them collaborate on common ground.

The middle portion of Figure~\ref{fig:personas} shows three
representative low-code techniques.
Programming expertise induces a Venn diagram over the users, with the
smallest subset being able to use the largest range of programming
techniques.
An edge between a set of users and a low-code technique indicates that
the users write or read a program with that technique.
Specifically, all users can use programming by demonstration and
programming by natural language~(edges to the outermost set of users
encompassing citizen-, semi-, and pro-developers).
Only semi-developers and pro-developers can readily use visual
programming, though citizen developers may be easily trained
to do so, as evidenced by Scratch~\cite{resnick_et_al_2009}.
And only pro-developers are likely to directly use
a domain-specific language~(DSL).
Therefore, while low-code typically targets a DSL, that DSL may not be
exposed, or if it is, may only be exposed to pro-developers.
That is especially true in the common case of a DSL that is
embedded~\cite{hudak_1998} in a general-purpose textual programming
language such as Python.

If the core value of low-code is to create computer programs, what
exactly is it about created programs that is deemed valuable?
One way to shed more light on this question is to look at a seemingly
opposing trend, namely the \emph{as-code} movement.
The as-code movement started with \emph{infrastructure as code}, which
automates standing up compute resources and the services running on
them from a source code repository and a backup~\cite{jacob_2010}.
By treating this process as code, organizations can speed it up,
reduce mistakes, and facilitate testing.
Another instance of as-code is \emph{security as code},
where security policies, templates, and configuration files all live
in a source code repository~\cite{myrbakken_colomopalacios_2017}.
By treating them as code, they can be versioned, inspected by
humans, and checked by machines.
To summarize, the as-code movement sees value in programs that are
repeatable, tested, versioned, human-readable, and machine-checkable.
These are also desirable properties for low-code programs.

When citizen developers use low-code, it is typically to create a
program that saves time on a task they would otherwise do by hand.
We will round out the problem statement by looking at
what tasks low-code is good for.
Generally speaking, low-code helps if it shaves off more time from
a task than the time spent doing the low-code programming.
This is true for tasks that are repetitive or time-consuming.
Of course, the equation shifts when the program can be used not just
by the developer who created it, but also by others, shaving time off
of their tasks as well.
In the extreme, pro-developers create programs used by millions.
Low-code is most appropriate when it saves time, but not enough time
to make professional coding economically feasible.
Low-code is suitable for tasks that are rule-based and low on
exceptions.
And besides the time savings, it can be even more beneficial when
the tedium of doing the task by hand causes errors.

\section{Techniques}\label{sec:techniques}

This section is a deep-dive into three representative techniques for
low-code programming:
VPLs~(visual programming languages), PBD~(programming by demonstration),
and PBNL~(programming by natural language).
These three are a good set for the following reasons.
Sahay et al.'s paper declares low-code as synonymous with just one
technique, VPLs~\cite{sahay_et_al_2020}, but we found that perspective
too narrow.
Barricelli et al.\ list 14 different techniques for end-user
programming~\cite{barricelli_et_al_2019}, but they are not clearly
separated, and reviewing them all in detail would get too long-winded.
In the past, the dominant low-code technique has been
spreadsheets~\cite{burnett_cook_rothermel_2004}.
The three techniques we chose instead align with present and future trends:
VPLs are central to current commercial low-code platforms~\cite{sahay_et_al_2020};
PBD is the back-bone of robotic process automation~(RPA), which often
uses record-and-replay~\cite{vanderaalst_bichler_heinzl_2018};
and PBNL is poised to grow thanks to advances in
deep learning based large language
models~\cite{chen_et_al_2021,shin_et_al_2021,yin_neubig_2017}.

Furthermore, the three techniques are well-suited for citizen
developers by drawing upon universal skills: VPLs draw upon seeing,
PBD draws upon the ability to use a computer application, and PBNL
draws upon speaking.
In fact, low-code can offer an alternative modality when some other
approach is impeded, such as using speech interfaces when a user's
hands or eyes are unavailable.
Finally, VPLs, PBD, and PBNL are sufficient to span a set of building
blocks that can also be arranged differently for use with other low-code
techniques, such as spreadsheets, rules, wizards, or templates.
Not all building blocks appear in all techniques, but the following
blocks recur enough to warrant brief up-front definitions:

\begin{itemize}
  \item \emph{code canvas:} renders code, e.g., visually as a flow
    graph
  \item \emph{palette:} offers components for drag-and-drop selection
  \item \emph{text box:} holds natural-language text used for
    code search, description, or generation
  \item \emph{player:} has buttons for capture, replay, pause, or step
  \item \emph{stage:} shows the effect of code execution
  \item \emph{configuration pane:} lets the user customize components,
    e.g., via graphical controls such as check-boxes or sliders, or
    textually by typing small formulas
\end{itemize}

Low-code techniques support not just writing programs, but also
reading and executing them.
Low-code techniques differ in which of the
above-listed building blocks are engaged to read, write, or execute
programs.
Whereas Figure~\ref{fig:personas} blurred the read/write/execute
distinction by using undirected edges, the rest of this section
explicates the distinction by using directed edges and colors
(\textcolor{read}{\textbf{orange}} for read, \textcolor{write}{\textbf{dark blue}} for write, and \textcolor{execute}{\textbf{purple}} for execute).

\subsection{Visual Programming Languages}\label{sec:vpl}

\begin{quote}\it
  The user drags visual components from a palette to a canvas,
  connects them, and configures them.
\end{quote}

\begin{figure}[!h]
\centerline{\includegraphics[width=.9\columnwidth]{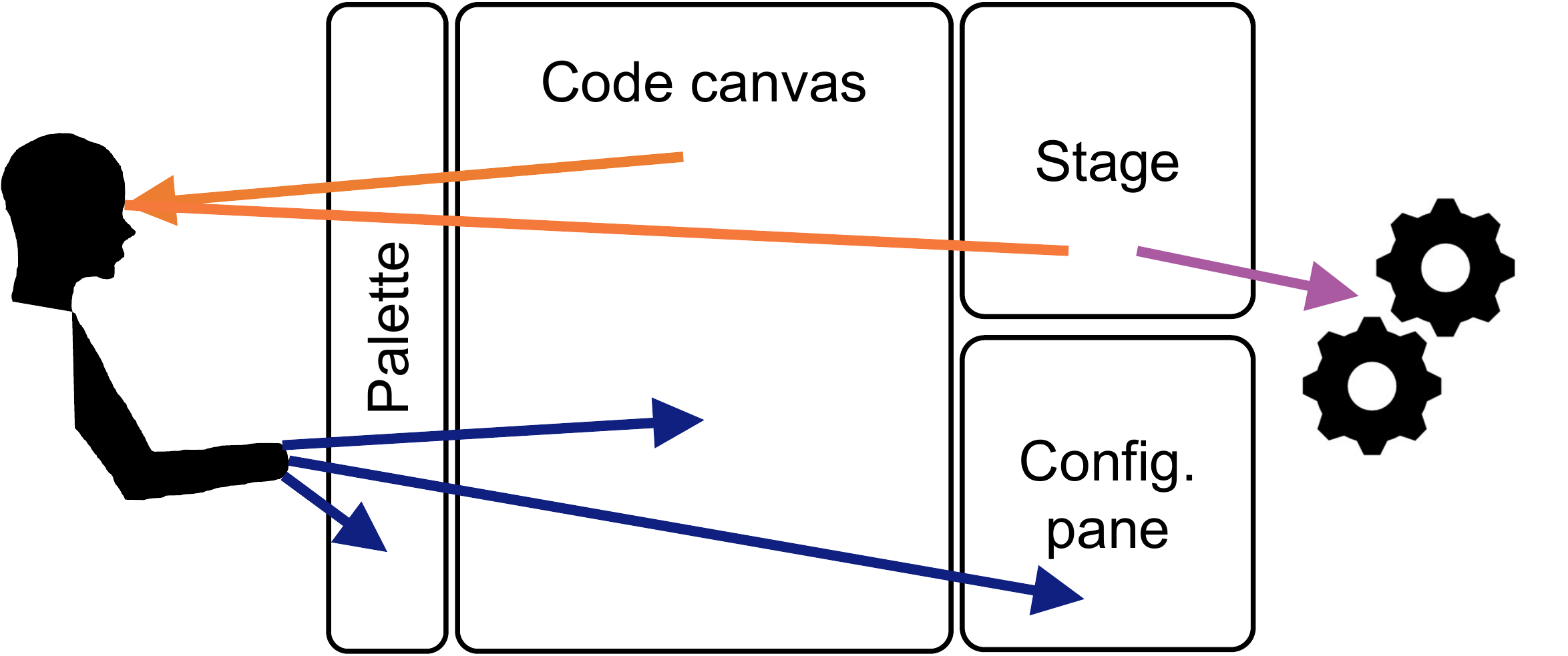}}
\caption{\label{fig:technique_vpl}Visual programming languages.}
\end{figure}

\subsubsection*{Description}

Visual programming languages let users write programs by directly
manipulating their visual representation.
There is a plethora of possible visual
representations~\cite{boshernitsan_downes_2004}, often inspired by
domain notation, such as electrical circuit diagrams.
Two prominent domain-independent visual representations are
boxes-and-arrows~(e.g., BPMN~\cite{ouyang_et_al_2006}) or interlocking
puzzle pieces~(e.g., Scratch~\cite{resnick_et_al_2009}).
Here, boxes or puzzle pieces represent instructions in the program,
and arrows between boxes or the interlock of pieces represent how data
and control flows between instructions.

Despite the diversity in visual languages, their programming
environments tend to comprise similar building blocks, as depicted in
Figure~\ref{fig:technique_vpl}.
The central building block is the code canvas, where the user can both
read~(red arrow from canvas to eye) and write~(blue arrow from hand
to canvas) the program.
Writing the program also involves dragging components from the palette
to the canvas and possibly configuring them in a separate
configuration pane.
The programming environment also often includes a stage, which
visually shows a concrete program execution in progress.
For example, in Scratch, the stage shows sprites in a virtual world.
Besides making the environment more engaging, the stage is also
crucial for program understanding and debugging.
To facilitate this, the stage is usually tightly connected to the
canvas, helping the user navigate back and forth.

\subsubsection*{Strengths, weaknesses, and mitigations}

One strength of VPLs is that they tend to be easy to read, either by
reusing notation that is already familiar to
the domain expert or by using a clean notation with general
appeal~\cite{boshernitsan_downes_2004}.
Another strength is that, in contrast to PBD or PBNL, VPLs are usually
unambiguous, thus increasing programmer control and reducing mistakes.
Finally, compared to textual programming languages,
visual languages can rule out syntax errors~\cite{voelter_lisson_2014}
and even simple type errors~\cite{resnick_et_al_2009} by construction.

In the context of low-code programming, the main weakness of visual
programming languages is that they are not always self-explanatory;
that is why Figure~\ref{fig:personas} connects them to semi-developers.
The mitigation for this need-to-learn is user education, and for some
VPLs, education is a primary purpose~\cite{resnick_et_al_2009}.
The visual notation can take up a lot of screen real estate; the
mitigation for this is to elide detail, e.g., by requiring a
configuration pane or via modular language
constructs~\cite{andersen_ballantyne_felleisen_2020,omar_et_al_2021}.
Even the palette can get too full, hindering discoverability, which
can be mitigated by search facilities.
A drawback of visual languages compared to textual languages is that
they tend to be co-dependent on their visual programming environment,
hindering the use of basic tools such as diffing or search, or of
third-party tools such as linters or code generators.
This can be mitigated by backing the visual language with a textual
domain-specific language~\cite{voelter_lisson_2014}.

\subsubsection*{Literature}

Some seminal VPLs include Harel's StateCharts system specification
language~\cite{harel_1987};
BPMN-on-BPEL for modeling and executing business
processes~\cite{ouyang_et_al_2006};
and the Scratch language for teaching kids
programming~\cite{resnick_et_al_2009}.
Boshernitsan and Downes chronicle early VPLs and categorize them
into purely visual vs.\ hybrid~(mixed with text), and complete
(sufficient procedural abstraction and data abstraction to be
self-hosting) or not~\cite{boshernitsan_downes_2004}.

Other papers address VPL implementation approaches, such as
\emph{meta-tools}~(tool used to implement other tools) and the
\emph{model-view-controller}~(MVC) pattern~(which lets users
manipulate the same model through multiple synchronized views).
VisPro is a meta-tool for creating visual programming
environments~\cite{zhang_zhang_cao_2001}.
VisPro advocates for a coordinated set of visual and textual
languages, using MVC to expose the same program~(model) via multiple
languages~(views).
More recently, Blockly is a meta-tool for creating VPLs with
interlocking puzzle pieces~\cite{pasternak_fenichel_marshall_2017}
like in Scratch.
And mage is a meta-tool for embedding VPLs in
notebooks~\cite{kery_et_al_2020}.
Some VPLs target pro-developers and are embedded in
professional programming environments or languages.
Projectional editing, such as in MPS~\cite{voelter_lisson_2014},
doubles down on the MVC paradigm, where even the textual language is
projected into a view that precludes syntax errors.
More recent work has demonstrated VPLs as libraries extending
textual languages such as
Racket~(a Lisp dialect)~\cite{andersen_ballantyne_felleisen_2020}
and Elm~(an ML dialect)~\cite{omar_et_al_2021}.

\subsection{Programming by Demonstration}\label{sec:pbd}

\begin{quote}\it
  The user demonstrates the behavior on a canvas, with some
  configuration during or after recording.
\end{quote}

\begin{figure}[!h]
\centerline{\includegraphics[width=.9\columnwidth]{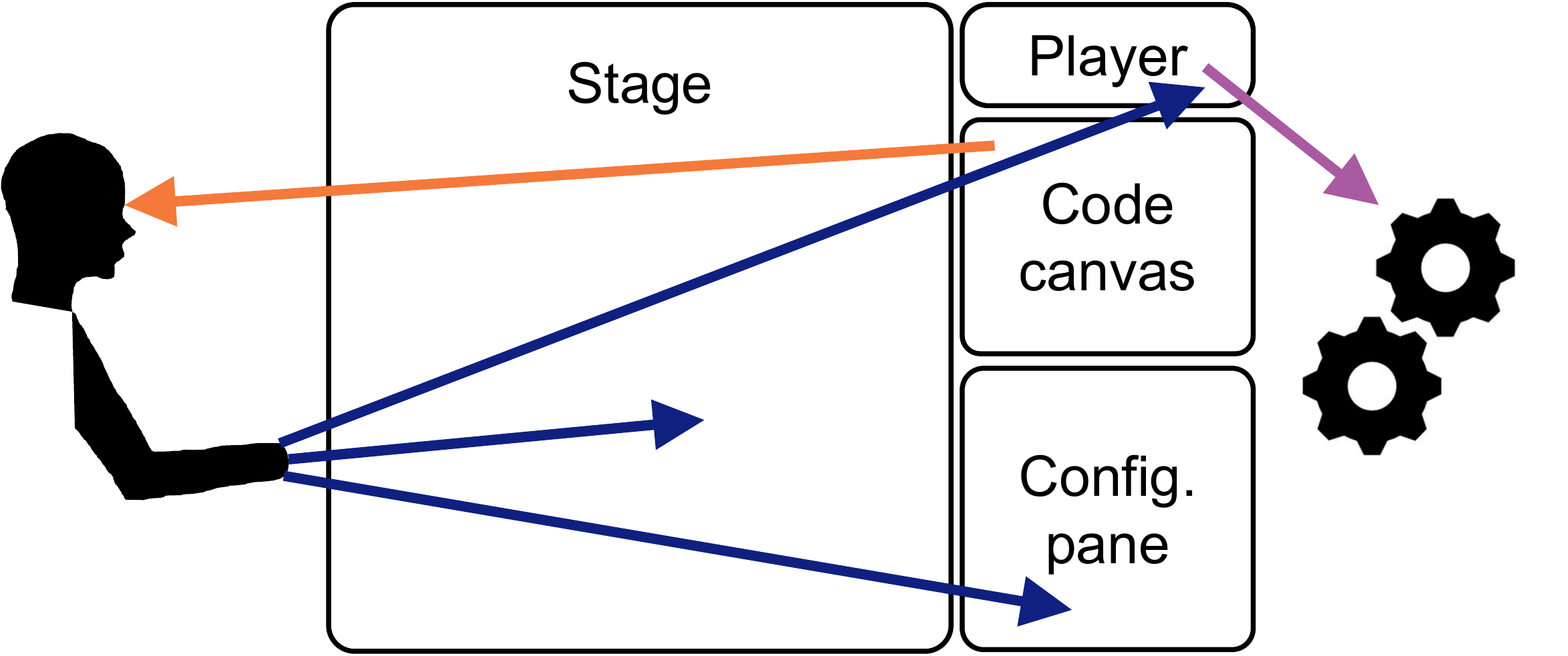}}
\caption{\label{fig:technique_pbd}Programming by demonstration.}
\end{figure}

\subsubsection*{Description}

In programming by demonstration, the user demonstrates how to perform
a task by hand via the mouse and keyboard, and the PBD system records
a program that can perform the same task automatically.
As shown in Figure~\ref{fig:technique_pbd}, the demonstration happens
on a stage, which may be a specific application like a spreadsheet, or
a web browser visiting a variety of sites and apps, or even a general
computer desktop or smart-phone screen.
Ideally, the recorded program abstracts from perceptions to a
symbolic representation, for instance, by mapping pixel coordinates to
a user-interface widget, or several keystrokes to a text string.
Besides the stage, most PBD systems have a player with
buttons to record and replay, plus often additional buttons such as
pause or step~(reminiscent of interactive debuggers).

The program is most useful if executing it does not yield exactly the
same behavior as the initial demonstration, but rather, generalizes to
different data.
For example, a program for ordering a taxi to any new location is more
general and more useful than a program for ordering a taxi to only a
single hard-coded location.
Generalizing typically requires identifying variables or parameters,
and may even entail adding conditionals, loops, or function calls.
Unfortunately, a single demonstration is an inherently ambiguous
specification for such a more general program.
Therefore, PBD systems often also provide a configuration pane that
allows users to disambiguate the generalization either during or after
demonstration.
Some PBD systems also have a code canvas that renders the recorded
program for the user to read, e.g., visually or in natural language.

\subsubsection*{Strengths, weaknesses, and mitigations}

The main strength of programming by demonstration is that the user can
work directly with the software applications they are already familiar
with from their day-to-day work~\cite{leshed_et_al_2008}.
This makes PBD well-suited for citizen developers, as there is no
indirection between programming and execution.
Furthermore, a demonstration is more concrete than a program in a
different paradigm, since it works on specific values and has a
straight-line flow of control and data.

Unfortunately, being so concrete is also PBD's main weakness: to turn
a demonstration into a program, it must be generalized, and automatic
generalization may not capture the user's intent~\cite{fischer_et_al_2021}.
Mitigations include hand-configuration~\cite{leshed_et_al_2008} or
multi-shot demonstration~\cite{gulwani_2011}.
PBD can be brittle with respect to the graphical user interface of the
application on stage, especially when that changes;
mitigations include heuristics and specialized
recorders that can map perception to application-level
concepts~\cite{sereshkeh_et_al_2020}.
Generalization can also overshoot, allowing a program to plow ahead
even in unforeseen circumstances~\cite{harel_marelly_2003}.
This can be mitigated by providing guard-rails, such as an attended
execution mode that asks the user to confirm before certain actions.
Finally, PBD can result in programs that are hard to understand
because they include spurious steps or are too fine-grained, which is
of course a problem in low-code programming~\cite{chasins_mueller_bodik_2018}.
This can be mitigated by pruning and by discovering macro-steps.

\subsubsection*{Literature}

A good example of a PBD system is CoScripter, where the stage is a web
browser and the code canvas displays the program in natural
language~\cite{leshed_et_al_2008}.
The CoScripter paper describes interviews that informed its design, as
well as experiences from real-world usage in a business setting.
In Rousillon, the stage is also a web browser and the canvas displays
the program in a VPL, fusing sequences of several low-level steps into
a single puzzle piece~\cite{chasins_mueller_bodik_2018}.
In VASTA, the stage is the display of a mobile phone, and the system
uses machine learning to reverse-engineer screenshots into user
interface elements~\cite{sereshkeh_et_al_2020}.
In DIYA, the stage is a web browser and users customize the program
during recording via voice input~\cite{fischer_et_al_2021}.
Robotic Process Automation applies PBD to business processes, by
letting a human business worker demonstrate a process on the existing
software and referring to the automatic replay engine as a
robot~\cite{vanderaalst_bichler_heinzl_2018}.

Programming by demonstration~(PBD) is closely related to programming
by example~(PBE), since a demonstration is an elaborate example.
FlashFill is a seminal PBE system that uses example input and output
columns in a spreadsheet to synthesize a program for transforming
inputs to outputs~\cite{gulwani_2011}.
Both PBD and PBE are based on program synthesis~\cite{alur_et_al_2018}.
Recent work has harnessed novel machine-learning techniques for
program synthesis, such as learned search strategies in
DeepCoder~\cite{balog_et_al_2017} and learned libraries in
DreamCoder~\cite{ellis_et_al_2021}.

PBD can be profitably combined with other low-code techniques.
The play-in / play-out approach is a PBD system co-designed with its
own VPL based on sequence diagrams~\cite{harel_marelly_2003}.
And SwaggerBot is a PBD system embedded in a natural-language
conversational agent, enabling a form of
PBNL~\cite{vaziri_et_al_2017}.

\subsection{Programming by Natural Language}\label{sec:pbnl}

\begin{quote}\it
  The user enters natural language text via keyboard or voice, and the
  system synthesizes a program.
\end{quote}

\begin{figure}[!h]
\centerline{\includegraphics[width=.9\columnwidth]{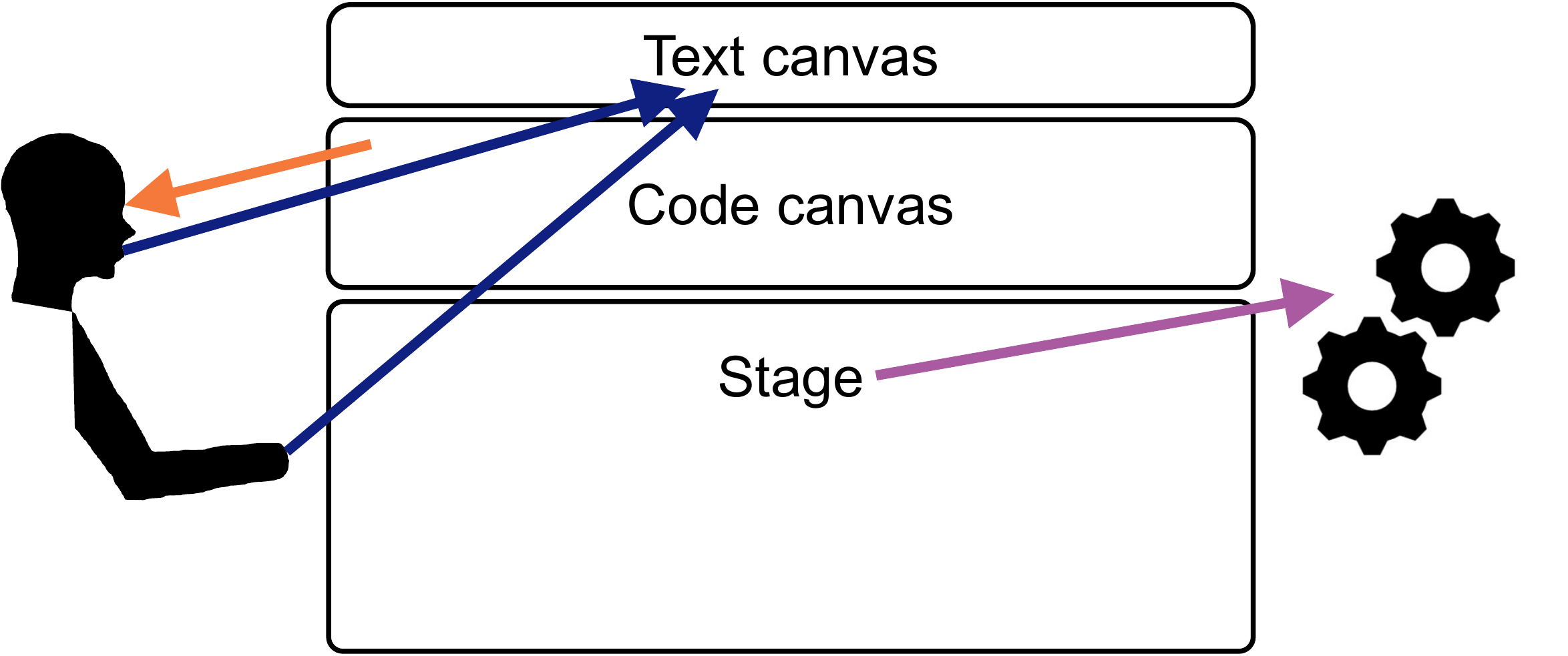}}
\caption{\label{fig:technique_pbnl}Programming by natural language.}
\end{figure}

\subsubsection*{Description}

In this low-code technique, the user enters text in natural language,
either by typing on the keyboard or via speech-to-text.
Figure~\ref{fig:technique_pbnl} indicates these two possibilities via
blue arrows from the user's hand or mouth to the text canvas.
The PBNL system translates the user's text, or utterance, to a program.
The system can optionally render the program on a code canvas for the
user to read.
This rendering might use a VPL, or it might use a controlled natural
language~\cite{kuhn_2014} for a disambiguated version of the
user's utterance.
The PBNL system can execute the program immediately or save it for
later, and the user may choose to execute the program multiple times,
e.g., after changes to the program's input data.
The system can also optionally show the effect of the program's execution
on a stage.
For example, if the program is a query in a spreadsheet, the
spreadsheet is the stage, and the result can be shown as a new table.

\subsubsection*{Strengths, weaknesses, and mitigations}

The main strength of PBNL is that it is not just low-code, but more
generally, low on demands during programming.
As shown in Figure~\ref{fig:technique_pbnl}, its programming
environment has only three building blocks~(text canvas, code canvas,
and stage), and all three are optional.
That means PBNL can in principle even be applied in circumstances
where the user's hands and eyes are otherwise occupied.
PBNL makes it particularly easy for citizen developers to create
programs.
Another strength of PBNL is its expressiveness: natural language can
express virtually anything humans want to communicate.
PBNL restricts neither the sophistication nor the domains of programs.

Unfortunately, while PBNL makes creating programs trivial, those
programs are often wrong~\cite{androutsopoulos_ritchie_thanisch_1995}.
Natural language is ambiguous, since humans are often vague
and tend to omit context or assume common ground.
On top of that, natural language processing~(NLP) technologies are
imperfect.
The optional code canvas and stage mitigate this
weakness, by showing the user the synthesized program or its effect,
thus giving them a chance to correct it.
Another mitigation is to encourage users to keep their utterances
short and not take advantage of the full expressiveness of natural
language, since simpler programs are easier to get right~\cite{li_et_al_2019}.
Furthermore, some PBNL systems support hand-editing the program.
Another weakness of PBNL systems is that they often require an aligned
corpus of utterances and programs to train machine-learning models,
and obtaining such a corpus is expensive.
Mitigating this is an active research topic in the machine-learning
research community~\cite{wang_berant_liang_2015,shin_et_al_2021}.

\subsubsection*{Literature}

As an interdisciplinary field of research, PBNL is best illuminated
through multiple surveys.
Androutsopoulos et al.\ surveyed natural-language interfaces to
databases, a prominent form of PBNL going back to the
1960s~\cite{androutsopoulos_ritchie_thanisch_1995}.
A common approach is to parse a natural-language utterance into a tree
and then map that tree to a database query.
Kuhn surveyed controlled natural languages~(CNLs), which restrict
inputs to be unambiguous while preserving some natural
properties~\cite{kuhn_2014}.
Compared to unrestricted natural language, CNLs may make it harder for
citizen developers to write programs, but may make it easier to write
correct programs.
Allamanis et al.\ surveyed machine learning for code, arguing that
code has a ``naturalness'' that makes it possible to adapt
various NLP technologies to work on code~\cite{allamanis_et_al_2018}.
The survey covers some code-generating models relevant to PBNL.

The most successful NLP technology applied to PBNL is semantic
parsers, which are machine-learning models that translate from natural
language to an abstract syntax tree~(AST) of a program.
For instance, SILT learns rule-based semantic parsers that have been
demonstrated for programs that coach robotic soccer teams or for
programs that query geographic databases~\cite{kate_wong_mooney_2005}.
The Overnight paper addresses the problem of obtaining an aligned
corpus for training a semantic parser via synthetic data generation
and crowd-sourced paraphrasing~\cite{wang_berant_liang_2015}.
Pumice tackles the ambiguity of natural language by a dialogue, where
the system prompts for clarification which the user can provide via
natural language or demonstration~\cite{li_et_al_2019}.
And Shin et al.\ show how to coax a pre-trained large
language model into doing semantic parsing without requiring
fine-tuning~\cite{shin_et_al_2021}.

Another approach to PBNL is program synthesis, which typically
searches a space of possible programs~\cite{alur_et_al_2018}.
Desai et al.\ describe a meta-synthesizer that, given a DSL grammar
and an aligned corpus, creates a synthesizer from natural language to
programs in the DSL~\cite{desai_et_al_2016}.
PBNL is not limited to domain-specific languages for citizen developers.
Yin and Neubig describe a semantic parser that uses deep learning to
encode a sequence of natural-language tokens, then decodes that into a
Python AST~\cite{yin_neubig_2017}.
Codex is a pre-trained large language model for natural language first
fine-tuned on unlabeled code, then fine-tuned again on an aligned
corpus of utterances and programs~\cite{chen_et_al_2021}.

\section{Perspectives}\label{sec:perspectives}

\begin{table}[!h]
\caption{\label{tab:techniques}Comparing low-code techniques.}
\centerline{\footnotesize\begin{tabular}{@{}cccccc@{}}
  \textbf{Technique} & \multicolumn{3}{c}{\textbf{Activity}} & \multirow{2}{*}{\parbox{8.5mm}{\centering\textbf{Ambi\-guity}}} & \multirow{2}{*}{\parbox{11mm}{\centering\vspace*{-.6mm}\textbf{Need to learn}}}\\
  \cmidrule(rl){2-4}
  & \textbf{\textcolor{write}{Write}} & \textbf{\textcolor{read}{Read}} & \textbf{\textcolor{execute}{Execute}}
  \\\midrule
  Visual        & code canvas, & code   & stage   & low    & medium\\
  programming   & palette,     & canvas &         &        & \\
  languages     & config.~pane &        &         &        &
  \\[2mm]
  Programming   & stage,       & code   & player, & medium & low\\
  by demon-     & player,      & canvas & stage   &        & \\
  stration      & config.~pane &        &         &        & 
  \\[2mm]
  Programming   & text canvas  & code   & stage   & high   & low\\
  by natural    &              & canvas &         &        & \\
  language      &              &        &         &        &
\end{tabular}}
\end{table}

While the previous section covered three low-code techniques in depth,
this section covers cross-cutting topics beyond any single technique.
Table~\ref{tab:techniques} compares the three techniques from
Section~\ref{sec:techniques}.
The Activity columns indicate how each technique supports the user in
writing, reading, and executing programs.
The main difference is in the Write column: users write programs
mainly on the code canvas for VPLs, the stage for PBD, and a text
canvas in PBNL.
On the other hand, there is little difference in the Read and Execute
columns: users read programs on a code canvas~(if provided), and
watch them executing on the stage~(if visible).
That hints at an opportunity for reusing building blocks across tools
for different techniques.

A core problem with low-code programming is ambiguity.
While visual programming languages can be rigorous and unambiguous,
there is ambiguity in how to generalize from a demonstration to a
program that works in different situations, and natural languages are
inherently ambiguous as well.
More ambiguous techniques may only work reliably on small and simple problems.
Systems for PBD and PBNL must guess at the user's intent, and are
likely to guess wrong when programs get complicated.
This motivates offering users an option to read or even correct
programs or their executions.

A core goal of low-code programming is to reduce the need to learn a
programming language.
Citizen developers can demonstrate a program or describe it in natural
language without having been taught how to do so.
Visual programming, on the other hand, is often not quite as
self-explanatory, which is why Figure~\ref{fig:personas} associates it
more with semi-developers.
On the other hand,
depending on the user's attitude, the need-to-learn can also be a
positive aspect, since it grows computational thinking skills.

\subsubsection*{Artificial Intelligence for Low-Code}

Does the ongoing rapid progress in AI fuel progress in low-code?
This article argues that yes, it does, in proportion to the ambiguity
of the low-code technique.
Out of the three techniques in Table~\ref{tab:techniques}, AI is most
prominent for PBNL, which is also the most ambiguous.
PBNL can hardly avoid AI except by using a controlled natural
language~\cite{kuhn_2014}, but that would make it feel more like code.
Currently a rising AI approach for PBNL is to use large language
models with code generation~\cite{chen_et_al_2021,shin_et_al_2021}.
We expect PBNL to grow along with relevant advances in AI.
AI is also prominent in PBD, which Table~\ref{tab:techniques}
characterizes as medium ambiguity.
For example, DeepCoder shows the interplay between program synthesis
for defining a space of possible programs and checking whether a given
program is correct, and AI for guiding the search through that
space~\cite{balog_et_al_2017}.
As another example, VASTA uses speech recognition, object recognition,
and optical character recognition to better understand a user's
demonstration of a task~\cite{sereshkeh_et_al_2020}.

\subsubsection*{Communicating with Humans and Machines}

Pro-developers use code in textual programming languages to
communicate with a computer, telling it what to do.
In addition, developers can also use programming languages to
communicate with each other or with their own future self.
A low-level programming language such as C gives developers more
control, whereas a high-level language
such as Python arguably makes communication among humans more effective.
Similarly, low-code programs can also serve both to communicate
instructions to a computer and to communicate among low-code users.
Being even more high-level than, say, Python, low-code can serve as a
lingua franca to help citizen developers and pro-developers
communicate more effectively with each other.
For instance, a citizen developer might use PBD to communicate a
desired behavior to a pro-developer to flesh out~\cite{harel_marelly_2003}.
Conversely, a pro-developer might use PBNL or a VPL to communicate a
proposed behavior to a domain expert for explanation or
approval~\cite{ouyang_et_al_2006}.

\subsubsection*{Domain-Specific Languages for Low-Code}

All three low-code techniques from Section~\ref{sec:techniques} are
intrinsically related to domain-specific languages~(DSLs):
most visual programming languages \emph{are} DSLs
(e.g.\ Scratch~\cite{resnick_et_al_2009}),
and both programming by demonstration and programming by natural
language usually \emph{target} DSLs~(e.g.\ DIYA targets its
co-designed ThingTalk 2.0 DSL~\cite{fischer_et_al_2021}).
Mernik et al.\ list further benefits of DSLs: they facilitate program
analysis, verification, optimization, parallelization, and
transformation~(AVOPT)~\cite{mernik_heering_sloane_2005}.

While reviewing the low-code literature reveals a close tie to DSLs,
those DSLs are not always exposed to the user.
For instance, the DSL may manifest as a proprietary file format or as
an undocumented internal representation.
If the DSL is exposed, users can more easily read, test, and audit
programs, version them and store them in a shared repository, and
manipulate them with tools for program transformation or generation.
Also, an exposed DSL is less locked into a specific programming
environment or its vendor.
When exposed, the DSL should be designed for humans, possibly based on
interviews and user studies as role-modeled by Leshed et al.~\cite{leshed_et_al_2008}.
On the other hand, a DSL that is not exposed will be shaped by
different factors, such as the ease of enumerating valid programs,
which can be improved by asymmetry~\cite{ellis_et_al_2021}.

DSLs~(including DSLs for low-code) may be embedded in a
general-purpose language.
Compared to a stand-alone DSL, an embedded DSL is often easier to
implement~(e.g., due to not requiring a custom parser) and easier to
use~(e.g., due to syntax highlighting and auto-completion tools of the
host language).
The approach to implementing an embedded DSL depends on the facilities
of the host language.
One approach is Pure Embedding, which uses higher-order functions and
lazy evaluation, such as in Haskell~\cite{hudak_1998}.
Another example is Lightweight Modular Staging, which uses
operator overloading and dynamic compilation, such as in
Scala~\cite{rompf_odersky_2012}.

\subsubsection*{Model View Controller}

\begin{figure}[!h]
\centerline{\includegraphics[width=\columnwidth]{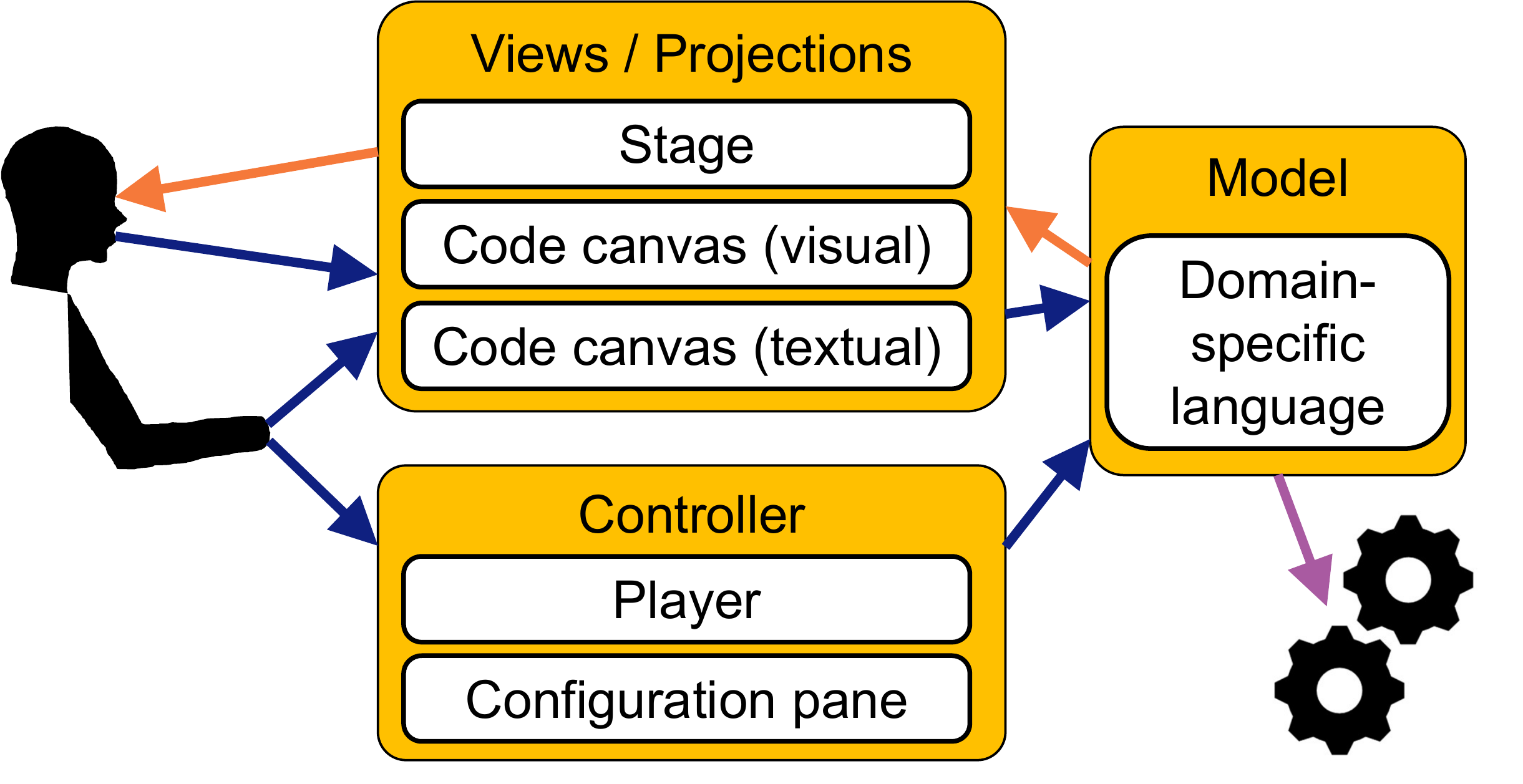}}
\caption{\label{fig:model_view_controller}Model-view-controller for low-code.}
\end{figure}

The current state-of-the-art VPLs and associated meta-tools are based
on the model-view-controller~(MVC) pattern.
And in PBD or PBNL, even though the user does not use a
code canvas to write a program, the system may optionally provide a
code canvas for reading it, in which case they also use MVC.
Figure~\ref{fig:model_view_controller} illustrates MVC with a superset
of the components from each low-code technique.
Low-code programming tools provide one or more views of the program.
Some of these views, or projections, are read-only, while others are
read-write views.
When multiple views are present, the system keeps them in synch with a
single joint model, and through that, with each other.
Edits in one view are projected live to all other views.
The model is a program in a domain-specific language~(DSL).
As discussed previously, the DSL may or may not be exposed to the
user, and may or may not be embedded in a host programming language.
Optionally, the system may even expose the textual DSL as another
view, for instance, in a structure editor~\cite{voelter_lisson_2014}.
Besides the model and the view, the third part of the MVC pattern is
the controller, which, for low-code, can contain a player and/or a
configuration pane.

\subsubsection*{Combining Multiple Low-Code Techniques}

When users write a program by demonstration or by natural language,
the system may let them read their program on a code canvas.
And once a system lets users read programs on a code canvas, a logical
next step is to also let them write programs there, such as, to
correct mistakes from generalization or from natural language processing.
This yields a combination of low-code techniques, where users can
write programs in multiple ways.
Such combinations can compensate for weaknesses of techniques.
For example, in Rousillon, the user first writes a program by
demonstrating how to scrape data from web
pages~\cite{chasins_mueller_bodik_2018}; since one weakness of PBD is
ambiguity, Rousillon next shows the resulting program to the user in a
scratch-like VPL~\cite{chasins_mueller_bodik_2018}.
As another example, Pumice combines PBD with PBNL: the user first
writes a program via natural language; since one weakness of PBNL is
ambiguity, Pumice next lets the user clarify with
PBD~\cite{li_et_al_2019}.

\subsubsection*{Meta-tools and Meta-circularity}

A \emph{meta-tool} for low-code is a tool that is used to implement
low-code tools.
In traditional programming languages, meta-tools~(such as parser
generators) have long been an essential part of the tool-writer's
repertoire.
Similarly, meta-tools for low-code can speed up the development of
low-code tools by automating well-known but tedious pieces.
Thus, meta-tools make it easier to build several tools or variants,
for instance, to experiment with the user experience.
There are examples of meta-tools for all three low-code techniques
discussed in Section~\ref{sec:techniques}:
Blockly is a VPL meta-tool~\cite{pasternak_fenichel_marshall_2017},
DreamCoder is a PBD meta-tool~\cite{ellis_et_al_2021}, and
Overnight is a PBNL meta-tool~\cite{wang_berant_liang_2015}.

A \emph{meta-circular tool} for low-code is a meta-tool for low-code
that is itself a low-code tool.
Not all meta-tools are meta-circular tools, as that requires them to
be powerful enough for serious software development.
Supporting all that power can compromise the tool's low-code nature:
complex features can get in the way of learning easy ones.
On the positive side, meta-circular tools can democratize the creation
of low-code tools themselves.
Furthermore, tool developers who use their own tools may
empathize more with their users' needs.
Examples for meta-circular low-code tools include
VisPRO~\cite{zhang_zhang_cao_2001} and
Racket~\cite{andersen_ballantyne_felleisen_2020}~(both for VPLs).

\subsubsection*{Low-Code Foundation}

In addition to meta-tools, are there other reusable modules that make
it easier to build new low-code tools?
The beginning of Section~\ref{sec:techniques} listed several reusable
building blocks for low-code programming interfaces: code canvas,
palette, text box, player, stage, and configuration pane.
Besides making it easier to create low-code tools, such reuse can also
give different tools a more uniform look-and-feel, thus reducing the
need-to-learn.
In the case of multiple low-code tools for the same domain, reusing
the same domain-specific language makes them more interoperable.
Of course, low-code tools in different domains will require different
DSLs, but they may still be able to reuse some sub-language, such as
expressions or formulas with basic arithmetic and logical operators
and a function library.
There are also AI components that can be reused across low-code tools,
such as speech recognition modules, a search-based program synthesis
engine, semantic parsers, or language models.

\subsubsection*{End-user Software Engineering}

Most of the discussion on low-code programming focuses on writing a
program: it enables citizen developers to rapidly create a prototype.
But what happens over time when these programs stick around, get used
in new circumstances that the developer did not foresee, get modified
or generalized, and proliferate?
At that point, users need end-user software engineering~(EUSE) for
quality control, for instance, by showing test coverage, letting users
add assertions, and helping them localize faults directly in their
low-code programming environment~\cite{burnett_cook_rothermel_2004}.
Another way to support EUSE is to expose the DSL, which makes it
easier to adopt established software development workflows and the
associated tools~(such as version-controlled source code repositories,
regression tests, or issue trackers) for low-code.

\section{Conclusion}\label{sec:conclusion}

This article reviews research relevant to low-code programming models
with a focus on visual programming, programming by demonstration, and
programming by natural language.
It maps low-code techniques to target users and discusses common
building blocks, strengths, and weaknesses.
This article argues that domain-specific languages and the
model-view-controller pattern constitute a common back-bone and
unifying principle across low-code techniques.

\newpage
\balance
\bibliography{bibfile}


\begin{thebibliography}{40}


\ifx \showCODEN    \undefined \def \showCODEN     #1{\unskip}     \fi
\ifx \showDOI      \undefined \def \showDOI       #1{#1}\fi
\ifx \showISBNx    \undefined \def \showISBNx     #1{\unskip}     \fi
\ifx \showISBNxiii \undefined \def \showISBNxiii  #1{\unskip}     \fi
\ifx \showISSN     \undefined \def \showISSN      #1{\unskip}     \fi
\ifx \showLCCN     \undefined \def \showLCCN      #1{\unskip}     \fi
\ifx \shownote     \undefined \def \shownote      #1{#1}          \fi
\ifx \showarticletitle \undefined \def \showarticletitle #1{#1}   \fi
\ifx \showURL      \undefined \def \showURL       {\relax}        \fi
\providecommand\bibfield[2]{#2}
\providecommand\bibinfo[2]{#2}
\providecommand\natexlab[1]{#1}
\providecommand\showeprint[2][]{arXiv:#2}

\bibitem[\protect\citeauthoryear{Allamanis, Barr, Devanbu, and
  Sutton}{Allamanis et~al\mbox{.}}{2018}]%
        {allamanis_et_al_2018}
\bibfield{author}{\bibinfo{person}{Miltiadis Allamanis},
  \bibinfo{person}{Earl~T Barr}, \bibinfo{person}{Premkumar Devanbu}, {and}
  \bibinfo{person}{Charles Sutton}.} \bibinfo{year}{2018}\natexlab{}.
\newblock \showarticletitle{A Survey of Machine Learning for Big Code and
  Naturalness}.
\newblock \bibinfo{journal}{\emph{ACM Computing Surveys (CSUR)}}
  \bibinfo{volume}{51}, \bibinfo{number}{4} (\bibinfo{date}{July}
  \bibinfo{year}{2018}), \bibinfo{pages}{81:1--81:37}.
\newblock
\urldef\tempurl%
\url{https://doi.org/10.1145/3212695}
\showURL{%
\tempurl}


\bibitem[\protect\citeauthoryear{Alur, Singh, Fisman, and Solar-Lezama}{Alur
  et~al\mbox{.}}{2018}]%
        {alur_et_al_2018}
\bibfield{author}{\bibinfo{person}{Rajeev Alur}, \bibinfo{person}{Rishabh
  Singh}, \bibinfo{person}{Dana Fisman}, {and} \bibinfo{person}{Armando
  Solar-Lezama}.} \bibinfo{year}{2018}\natexlab{}.
\newblock \showarticletitle{Search-Based Program Synthesis}.
\newblock \bibinfo{journal}{\emph{Communications of the ACM (CACM)}}
  (\bibinfo{date}{Nov.} \bibinfo{year}{2018}), \bibinfo{pages}{84--93}.
\newblock
\urldef\tempurl%
\url{https://doi.org/10.1145/3208071}
\showURL{%
\tempurl}


\bibitem[\protect\citeauthoryear{Andersen, Ballantyne, and Felleisen}{Andersen
  et~al\mbox{.}}{2020}]%
        {andersen_ballantyne_felleisen_2020}
\bibfield{author}{\bibinfo{person}{Leif Andersen}, \bibinfo{person}{Michael
  Ballantyne}, {and} \bibinfo{person}{Matthias Felleisen}.}
  \bibinfo{year}{2020}\natexlab{}.
\newblock \showarticletitle{Adding Interactive Visual Syntax to Textual Code}.
  In \bibinfo{booktitle}{\emph{Conference on Object-Oriented Programming,
  Systems, Languages, and Applications (OOPSLA)}}.
\newblock
\urldef\tempurl%
\url{https://doi.org/10.1145/3428290}
\showURL{%
\tempurl}


\bibitem[\protect\citeauthoryear{Androutsopoulos, Ritchie, and
  Thanisch}{Androutsopoulos et~al\mbox{.}}{1995}]%
        {androutsopoulos_ritchie_thanisch_1995}
\bibfield{author}{\bibinfo{person}{Ion Androutsopoulos},
  \bibinfo{person}{Graeme~D. Ritchie}, {and} \bibinfo{person}{Peter Thanisch}.}
  \bibinfo{year}{1995}\natexlab{}.
\newblock \showarticletitle{Natural Language Interfaces to Databases -- An
  Introduction}.
\newblock \bibinfo{journal}{\emph{Natural Language Engineering}}
  \bibinfo{volume}{1}, \bibinfo{number}{1} (\bibinfo{year}{1995}),
  \bibinfo{pages}{29--81}.
\newblock
\urldef\tempurl%
\url{https://doi.org/10.1017/S135132490000005X}
\showURL{%
\tempurl}


\bibitem[\protect\citeauthoryear{Balog, Gaunt, Brockschmidt, Nowozin, and
  Tarlow}{Balog et~al\mbox{.}}{2017}]%
        {balog_et_al_2017}
\bibfield{author}{\bibinfo{person}{Matej Balog}, \bibinfo{person}{Alexander~L.
  Gaunt}, \bibinfo{person}{Marc Brockschmidt}, \bibinfo{person}{Sebastian
  Nowozin}, {and} \bibinfo{person}{Daniel Tarlow}.}
  \bibinfo{year}{2017}\natexlab{}.
\newblock \showarticletitle{{DeepCoder}: Learning to Write Programs}. In
  \bibinfo{booktitle}{\emph{International Conference on Learning
  Representations (ICLR)}}.
\newblock
\urldef\tempurl%
\url{https://openreview.net/forum?id=ByldLrqlx}
\showURL{%
\tempurl}


\bibitem[\protect\citeauthoryear{Barricelli, Cassano, Fogli, and
  Piccinno}{Barricelli et~al\mbox{.}}{2019}]%
        {barricelli_et_al_2019}
\bibfield{author}{\bibinfo{person}{Barbara~Rita Barricelli},
  \bibinfo{person}{Fabio Cassano}, \bibinfo{person}{Daniela Fogli}, {and}
  \bibinfo{person}{Antonio Piccinno}.} \bibinfo{year}{2019}\natexlab{}.
\newblock \showarticletitle{End-user development, end-user programming and
  end-user software engineering: A systematic mapping study}.
\newblock \bibinfo{journal}{\emph{Journal of Systems and Software}}
  \bibinfo{volume}{149} (\bibinfo{year}{2019}), \bibinfo{pages}{101--137}.
\newblock
\urldef\tempurl%
\url{https://doi.org/10.1016/j.jss.2018.11.041}
\showURL{%
\tempurl}


\bibitem[\protect\citeauthoryear{Bock and Frank}{Bock and Frank}{2021}]%
        {bock_frank_2021}
\bibfield{author}{\bibinfo{person}{Alexander~C. Bock} {and}
  \bibinfo{person}{Ulrich Frank}.} \bibinfo{year}{2021}\natexlab{}.
\newblock \showarticletitle{Low-Code Platform}.
\newblock \bibinfo{journal}{\emph{Business \& Information Systems Engineering
  (BISE)}}  \bibinfo{volume}{63} (\bibinfo{year}{2021}),
  \bibinfo{pages}{733--740}.
\newblock
\urldef\tempurl%
\url{https://doi.org/10.1007/s12599-021-00726-8}
\showURL{%
\tempurl}


\bibitem[\protect\citeauthoryear{Boshernitsan and Downes}{Boshernitsan and
  Downes}{2004}]%
        {boshernitsan_downes_2004}
\bibfield{author}{\bibinfo{person}{Marat Boshernitsan} {and}
  \bibinfo{person}{Michael Downes}.} \bibinfo{year}{2004}\natexlab{}.
\newblock \bibinfo{booktitle}{\emph{Visual Programming Languages: A Survey}}.
\newblock \bibinfo{type}{{T}echnical {R}eport} UCB/CSD-04-1368.
  \bibinfo{institution}{University of California, Berkeley}.
\newblock
\urldef\tempurl%
\url{https://digitalassets.lib.berkeley.edu/techreports/ucb/text/CSD-04-1368.pdf}
\showURL{%
\tempurl}


\bibitem[\protect\citeauthoryear{Burnett, Cook, and Rothermel}{Burnett
  et~al\mbox{.}}{2004}]%
        {burnett_cook_rothermel_2004}
\bibfield{author}{\bibinfo{person}{Margaret Burnett}, \bibinfo{person}{Curtis
  Cook}, {and} \bibinfo{person}{Gregg Rothermel}.}
  \bibinfo{year}{2004}\natexlab{}.
\newblock \showarticletitle{End-User Software Engineering}.
\newblock \bibinfo{journal}{\emph{Communications of the ACM (CACM)}}
  \bibinfo{volume}{47}, \bibinfo{number}{9} (\bibinfo{date}{Sept.}
  \bibinfo{year}{2004}), \bibinfo{pages}{53--58}.
\newblock
\urldef\tempurl%
\url{https://doi.org/10.1145/1015864.1015889}
\showURL{%
\tempurl}


\bibitem[\protect\citeauthoryear{Chasins, Mueller, and Bodik}{Chasins
  et~al\mbox{.}}{2018}]%
        {chasins_mueller_bodik_2018}
\bibfield{author}{\bibinfo{person}{Sarah~E. Chasins}, \bibinfo{person}{Maria
  Mueller}, {and} \bibinfo{person}{Rastislav Bodik}.}
  \bibinfo{year}{2018}\natexlab{}.
\newblock \showarticletitle{Rousillon: Scraping Distributed Hierarchical Web
  Data}. In \bibinfo{booktitle}{\emph{Symposium on User Interface Software and
  Technology (UIST)}}. \bibinfo{pages}{963--975}.
\newblock
\urldef\tempurl%
\url{https://doi.org/10.1145/3242587.3242661}
\showURL{%
\tempurl}


\bibitem[\protect\citeauthoryear{Chen, Tworek, Jun, Yuan, Ponde, Kaplan,
  Edwards, Burda, Joseph, Brockman, Ray, Puri, Krueger, Petrov, Khlaaf, Sastry,
  Mishkin, Chan, Gray, Ryder, Pavlov, Power, Kaiser, Bavarian, Winter, Tillet,
  Such, Cummings, Plappert, Chantzis, Barnes, Herbert-Voss, Guss, Nichol,
  Babuschkin, Balaji, Jain, Carr, Leike, Achiam, Misra, Morikawa, Radford,
  Knight, Brundage, Murati, Mayer, Welinder, McGrew, Amodei, McCandlish,
  Sutskever, and Zaremba}{Chen et~al\mbox{.}}{2021}]%
        {chen_et_al_2021}
\bibfield{author}{\bibinfo{person}{Mark Chen}, \bibinfo{person}{Jerry Tworek},
  \bibinfo{person}{Heewoo Jun}, \bibinfo{person}{Qiming Yuan},
  \bibinfo{person}{Henrique Ponde}, \bibinfo{person}{Jared Kaplan},
  \bibinfo{person}{Harri Edwards}, \bibinfo{person}{Yura Burda},
  \bibinfo{person}{Nicholas Joseph}, \bibinfo{person}{Greg Brockman},
  \bibinfo{person}{Alex Ray}, \bibinfo{person}{Raul Puri},
  \bibinfo{person}{Gretchen Krueger}, \bibinfo{person}{Michael Petrov},
  \bibinfo{person}{Heidy Khlaaf}, \bibinfo{person}{Girish Sastry},
  \bibinfo{person}{Pamela Mishkin}, \bibinfo{person}{Brooke Chan},
  \bibinfo{person}{Scott Gray}, \bibinfo{person}{Nick Ryder},
  \bibinfo{person}{Mikhail Pavlov}, \bibinfo{person}{Alethea Power},
  \bibinfo{person}{Lukasz Kaiser}, \bibinfo{person}{Mohammad Bavarian},
  \bibinfo{person}{Clemens Winter}, \bibinfo{person}{Philippe Tillet},
  \bibinfo{person}{Felipe Such}, \bibinfo{person}{Dave Cummings},
  \bibinfo{person}{Matthias Plappert}, \bibinfo{person}{Fotios Chantzis},
  \bibinfo{person}{Elizabeth Barnes}, \bibinfo{person}{Ariel Herbert-Voss},
  \bibinfo{person}{Will Guss}, \bibinfo{person}{Alex Nichol},
  \bibinfo{person}{Igor Babuschkin}, \bibinfo{person}{Suchir Balaji},
  \bibinfo{person}{Shantanu Jain}, \bibinfo{person}{Andrew Carr},
  \bibinfo{person}{Jan Leike}, \bibinfo{person}{Josh Achiam},
  \bibinfo{person}{Vedant Misra}, \bibinfo{person}{Evan Morikawa},
  \bibinfo{person}{Alec Radford}, \bibinfo{person}{Matthew Knight},
  \bibinfo{person}{Miles Brundage}, \bibinfo{person}{Mira Murati},
  \bibinfo{person}{Katie Mayer}, \bibinfo{person}{Peter Welinder},
  \bibinfo{person}{Bob McGrew}, \bibinfo{person}{Dario Amodei},
  \bibinfo{person}{Sam McCandlish}, \bibinfo{person}{Ilya Sutskever}, {and}
  \bibinfo{person}{Wojciech Zaremba}.} \bibinfo{year}{2021}\natexlab{}.
\newblock \bibinfo{title}{Evaluating Large Language Models Trained on Code}.
\newblock
\newblock
\urldef\tempurl%
\url{https://arxiv.org/abs/2107.03374}
\showURL{%
\tempurl}


\bibitem[\protect\citeauthoryear{Desai, Gulwani, Hingorani, Jain, Karkare,
  Marron, R, and Roy}{Desai et~al\mbox{.}}{2016}]%
        {desai_et_al_2016}
\bibfield{author}{\bibinfo{person}{Aditya Desai}, \bibinfo{person}{Sumit
  Gulwani}, \bibinfo{person}{Vineet Hingorani}, \bibinfo{person}{Nidhi Jain},
  \bibinfo{person}{Amey Karkare}, \bibinfo{person}{Mark Marron},
  \bibinfo{person}{Sailesh R}, {and} \bibinfo{person}{Subhajit Roy}.}
  \bibinfo{year}{2016}\natexlab{}.
\newblock \showarticletitle{Program Synthesis Using Natural Language}. In
  \bibinfo{booktitle}{\emph{International Conference on Software Engineering
  (ICSE)}}. \bibinfo{pages}{345--356}.
\newblock
\urldef\tempurl%
\url{https://doi.org/10.1145/2884781.2884786}
\showURL{%
\tempurl}


\bibitem[\protect\citeauthoryear{Ellis, Wong, Nye, Sabl\'{e}-Meyer, Morales,
  Hewitt, Cary, Solar-Lezama, and Tenenbaum}{Ellis et~al\mbox{.}}{2021}]%
        {ellis_et_al_2021}
\bibfield{author}{\bibinfo{person}{Kevin Ellis}, \bibinfo{person}{Catherine
  Wong}, \bibinfo{person}{Maxwell Nye}, \bibinfo{person}{Mathias
  Sabl\'{e}-Meyer}, \bibinfo{person}{Lucas Morales}, \bibinfo{person}{Luke
  Hewitt}, \bibinfo{person}{Luc Cary}, \bibinfo{person}{Armando Solar-Lezama},
  {and} \bibinfo{person}{Joshua~B. Tenenbaum}.}
  \bibinfo{year}{2021}\natexlab{}.
\newblock \showarticletitle{{DreamCoder}: Bootstrapping Inductive Program
  Synthesis with Wake-Sleep Library Learning}. In
  \bibinfo{booktitle}{\emph{Conference on Programming Language Design and
  Implementation (PLDI)}}. \bibinfo{pages}{835--850}.
\newblock
\urldef\tempurl%
\url{https://doi.org/10.1145/3453483.3454080}
\showURL{%
\tempurl}


\bibitem[\protect\citeauthoryear{Fischer, Campagna, Choi, and Lam}{Fischer
  et~al\mbox{.}}{2021}]%
        {fischer_et_al_2021}
\bibfield{author}{\bibinfo{person}{Michael~H. Fischer},
  \bibinfo{person}{Giovanni Campagna}, \bibinfo{person}{Euirim Choi}, {and}
  \bibinfo{person}{Monica~S. Lam}.} \bibinfo{year}{2021}\natexlab{}.
\newblock \showarticletitle{{DIY} Assistant: A Multi-Modal End-User
  Programmable Virtual Assistant}. In \bibinfo{booktitle}{\emph{Conference on
  Programming Language Design and Implementation (PLDI)}}.
  \bibinfo{pages}{312--327}.
\newblock
\urldef\tempurl%
\url{https://doi.org/10.1145/3453483.3454046}
\showURL{%
\tempurl}


\bibitem[\protect\citeauthoryear{Gulwani}{Gulwani}{2011}]%
        {gulwani_2011}
\bibfield{author}{\bibinfo{person}{Sumit Gulwani}.}
  \bibinfo{year}{2011}\natexlab{}.
\newblock \showarticletitle{Automating String Processing in Spreadsheets using
  Input-Output Examples}. In \bibinfo{booktitle}{\emph{Symposium on Principles
  of Programming Languages (POPL)}}. \bibinfo{pages}{317--330}.
\newblock
\urldef\tempurl%
\url{https://doi.org/10.1145/1926385.1926423}
\showURL{%
\tempurl}


\bibitem[\protect\citeauthoryear{Harel}{Harel}{1987}]%
        {harel_1987}
\bibfield{author}{\bibinfo{person}{David Harel}.}
  \bibinfo{year}{1987}\natexlab{}.
\newblock \showarticletitle{{StateCharts}: a Visual Formalism for Complex
  Systems}.
\newblock \bibinfo{journal}{\emph{Science of Computer Programming}}
  \bibinfo{volume}{8}, \bibinfo{number}{3} (\bibinfo{year}{1987}),
  \bibinfo{pages}{231--274}.
\newblock
\urldef\tempurl%
\url{https://doi.org/10.1016/0167-6423(87)90035-9}
\showURL{%
\tempurl}


\bibitem[\protect\citeauthoryear{Harel and Marelly}{Harel and Marelly}{2003}]%
        {harel_marelly_2003}
\bibfield{author}{\bibinfo{person}{David Harel} {and} \bibinfo{person}{Rami
  Marelly}.} \bibinfo{year}{2003}\natexlab{}.
\newblock \showarticletitle{Specifying and executing behavioral requirements:
  the play-in/play-out approach}.
\newblock \bibinfo{journal}{\emph{Software and Systems Modeling (SoSyM)}}
  \bibinfo{volume}{2} (\bibinfo{year}{2003}), \bibinfo{pages}{82--107}.
\newblock
\urldef\tempurl%
\url{https://doi.org/10.1007/s10270-002-0015-5}
\showURL{%
\tempurl}


\bibitem[\protect\citeauthoryear{Hudak}{Hudak}{1998}]%
        {hudak_1998}
\bibfield{author}{\bibinfo{person}{Paul Hudak}.}
  \bibinfo{year}{1998}\natexlab{}.
\newblock \showarticletitle{Modular domain specific languages and tools}. In
  \bibinfo{booktitle}{\emph{International Conference on Software Reuse
  (ICSR)}}. \bibinfo{pages}{134--142}.
\newblock
\urldef\tempurl%
\url{https://doi.org/10.1109/ICSR.1998.685738}
\showURL{%
\tempurl}


\bibitem[\protect\citeauthoryear{Jacob}{Jacob}{2010}]%
        {jacob_2010}
\bibfield{author}{\bibinfo{person}{Adam Jacob}.}
  \bibinfo{year}{2010}\natexlab{}.
\newblock \showarticletitle{Infrastructure as Code}.
\newblock In \bibinfo{booktitle}{\emph{Web Operations: Keeping the Data on
  Time}}, \bibfield{editor}{\bibinfo{person}{John Allspaw} {and}
  \bibinfo{person}{Jesse Robbins}} (Eds.). \bibinfo{publisher}{O'Reilly},
  Chapter~5, \bibinfo{pages}{65--80}.
\newblock


\bibitem[\protect\citeauthoryear{Kate, Wong, and Mooney}{Kate
  et~al\mbox{.}}{2005}]%
        {kate_wong_mooney_2005}
\bibfield{author}{\bibinfo{person}{Rohit~J. Kate}, \bibinfo{person}{Yuk~Wah
  Wong}, {and} \bibinfo{person}{Raymond~J. Mooney}.}
  \bibinfo{year}{2005}\natexlab{}.
\newblock \showarticletitle{Learning to Transform Natural to Formal Languages}.
  In \bibinfo{booktitle}{\emph{Conference on Artificial Intelligence (AAAI)}}.
  \bibinfo{pages}{1062--1068}.
\newblock
\urldef\tempurl%
\url{http://www.aaai.org/Library/AAAI/2005/aaai05-168.php}
\showURL{%
\tempurl}


\bibitem[\protect\citeauthoryear{Kery, Ren, Hohman, Moritz, Wongsuphasawat, and
  Patel}{Kery et~al\mbox{.}}{2020}]%
        {kery_et_al_2020}
\bibfield{author}{\bibinfo{person}{Mary~Beth Kery}, \bibinfo{person}{Donghao
  Ren}, \bibinfo{person}{Fred Hohman}, \bibinfo{person}{Dominik Moritz},
  \bibinfo{person}{Kanit Wongsuphasawat}, {and} \bibinfo{person}{Kayur Patel}.}
  \bibinfo{year}{2020}\natexlab{}.
\newblock \showarticletitle{mage: Fluid Moves Between Code and Graphical Work
  in Computational Notebooks}. In \bibinfo{booktitle}{\emph{Symposium on User
  Interface Software and Technology (UIST)}}. \bibinfo{pages}{140--151}.
\newblock
\urldef\tempurl%
\url{https://doi.org/10.1145/3379337.3415842}
\showURL{%
\tempurl}


\bibitem[\protect\citeauthoryear{Kuhn}{Kuhn}{2014}]%
        {kuhn_2014}
\bibfield{author}{\bibinfo{person}{Tobias Kuhn}.}
  \bibinfo{year}{2014}\natexlab{}.
\newblock \showarticletitle{A Survey and Classification of Controlled Natural
  Languages}.
\newblock \bibinfo{journal}{\emph{Computational Linguistics}}
  \bibinfo{volume}{40}, \bibinfo{number}{1} (\bibinfo{year}{2014}),
  \bibinfo{pages}{121--170}.
\newblock
\urldef\tempurl%
\url{http://www.mitpressjournals.org/doi/pdf/10.1162/COLI_a_00168}
\showURL{%
\tempurl}


\bibitem[\protect\citeauthoryear{Leshed, Haber, Matthews, and Lau}{Leshed
  et~al\mbox{.}}{2008}]%
        {leshed_et_al_2008}
\bibfield{author}{\bibinfo{person}{Gilly Leshed}, \bibinfo{person}{Eben~M.
  Haber}, \bibinfo{person}{Tara Matthews}, {and} \bibinfo{person}{Tessa Lau}.}
  \bibinfo{year}{2008}\natexlab{}.
\newblock \showarticletitle{{CoScripter}: Automating \& Sharing How-to
  Knowledge in the Enterprise}. In \bibinfo{booktitle}{\emph{Conference on
  Human Factors in Computing Systems (CHI)}}. \bibinfo{pages}{1719--1728}.
\newblock
\urldef\tempurl%
\url{https://doi.org/10.1145/1357054.1357323}
\showURL{%
\tempurl}


\bibitem[\protect\citeauthoryear{Li, Radensky, Jia, Singarajah, Mitchell, and
  Myers}{Li et~al\mbox{.}}{2019}]%
        {li_et_al_2019}
\bibfield{author}{\bibinfo{person}{Toby Jia-Jun Li}, \bibinfo{person}{Marissa
  Radensky}, \bibinfo{person}{Justin Jia}, \bibinfo{person}{Kirielle
  Singarajah}, \bibinfo{person}{Tom~M. Mitchell}, {and}
  \bibinfo{person}{Brad~A. Myers}.} \bibinfo{year}{2019}\natexlab{}.
\newblock \showarticletitle{{PUMICE}: A Multi-Modal Agent That Learns Concepts
  and Conditionals from Natural Language and Demonstrations}. In
  \bibinfo{booktitle}{\emph{Symposium on User Interface Software and Technology
  (UIST)}}. \bibinfo{pages}{577--589}.
\newblock
\urldef\tempurl%
\url{https://doi.org/10.1145/3332165.3347899}
\showURL{%
\tempurl}


\bibitem[\protect\citeauthoryear{Mernik, Heering, and Sloane}{Mernik
  et~al\mbox{.}}{2005}]%
        {mernik_heering_sloane_2005}
\bibfield{author}{\bibinfo{person}{Marjan Mernik}, \bibinfo{person}{Jan
  Heering}, {and} \bibinfo{person}{Anthony~M. Sloane}.}
  \bibinfo{year}{2005}\natexlab{}.
\newblock \showarticletitle{When and how to develop domain-specific languages}.
\newblock \bibinfo{journal}{\emph{ACM Computing Surveys (CSUR)}}
  \bibinfo{volume}{37}, \bibinfo{number}{4} (\bibinfo{year}{2005}),
  \bibinfo{pages}{316--344}.
\newblock
\urldef\tempurl%
\url{https://doi.org/10.1145/1118890.1118892}
\showURL{%
\tempurl}


\bibitem[\protect\citeauthoryear{Myrbakken and Colomo-Palacios}{Myrbakken and
  Colomo-Palacios}{2017}]%
        {myrbakken_colomopalacios_2017}
\bibfield{author}{\bibinfo{person}{Havard Myrbakken} {and}
  \bibinfo{person}{Ricardo Colomo-Palacios}.} \bibinfo{year}{2017}\natexlab{}.
\newblock \showarticletitle{{DevSecOps}: A Multivocal Literature Review}. In
  \bibinfo{booktitle}{\emph{Software Process Improvement and Capability
  Determination (SPICE)}}. \bibinfo{pages}{17--29}.
\newblock
\urldef\tempurl%
\url{https://doi.org/10.1007/978-3-319-67383-7_2}
\showURL{%
\tempurl}


\bibitem[\protect\citeauthoryear{Omar, Moon, Blinn, Voysey, Collins, and
  Chugh}{Omar et~al\mbox{.}}{2021}]%
        {omar_et_al_2021}
\bibfield{author}{\bibinfo{person}{Cyrus Omar}, \bibinfo{person}{David Moon},
  \bibinfo{person}{Andrew Blinn}, \bibinfo{person}{Ian Voysey},
  \bibinfo{person}{Nick Collins}, {and} \bibinfo{person}{Ravi Chugh}.}
  \bibinfo{year}{2021}\natexlab{}.
\newblock \showarticletitle{Filling Typed Holes with Live {GUIs}}. In
  \bibinfo{booktitle}{\emph{Conference on Programming Language Design and
  Implementation (PLDI)}}. \bibinfo{pages}{511--525}.
\newblock
\urldef\tempurl%
\url{https://doi.org/10.1145/3453483.3454059}
\showURL{%
\tempurl}


\bibitem[\protect\citeauthoryear{Ouyang, Dumas, {Ter Hofstede}, and {Van Der
  Aalst}}{Ouyang et~al\mbox{.}}{2006}]%
        {ouyang_et_al_2006}
\bibfield{author}{\bibinfo{person}{Chun Ouyang}, \bibinfo{person}{Marlon
  Dumas}, \bibinfo{person}{Arthur~H.M. {Ter Hofstede}}, {and}
  \bibinfo{person}{Wil~M.P. {Van Der Aalst}}.} \bibinfo{year}{2006}\natexlab{}.
\newblock \showarticletitle{From {BPMN} Process Models to {BPEL} Web Services}.
  In \bibinfo{booktitle}{\emph{International Conference on Web Services
  (ICWS)}}.
\newblock
\urldef\tempurl%
\url{https://doi.org/10.1109/ICWS.2006.67}
\showURL{%
\tempurl}


\bibitem[\protect\citeauthoryear{Pasternak, Fenichel, and Marshall}{Pasternak
  et~al\mbox{.}}{2017}]%
        {pasternak_fenichel_marshall_2017}
\bibfield{author}{\bibinfo{person}{Erik Pasternak}, \bibinfo{person}{Rachel
  Fenichel}, {and} \bibinfo{person}{Andrew~N. Marshall}.}
  \bibinfo{year}{2017}\natexlab{}.
\newblock \showarticletitle{Tips for Creating a Block Language with {Blockly}}.
  In \bibinfo{booktitle}{\emph{Blocks and Beyond Workshop (B\&B)}}.
\newblock
\urldef\tempurl%
\url{https://doi.org/10.1109/BLOCKS.2017.8120404}
\showURL{%
\tempurl}


\bibitem[\protect\citeauthoryear{Resnick, Maloney, Monroy-Hern\'{a}ndez, Rusk,
  Eastmond, Brennan, Millner, Rosenbaum, Silver, Silverman, and Kafai}{Resnick
  et~al\mbox{.}}{2009}]%
        {resnick_et_al_2009}
\bibfield{author}{\bibinfo{person}{Mitchel Resnick}, \bibinfo{person}{John
  Maloney}, \bibinfo{person}{Andr\'{e}s Monroy-Hern\'{a}ndez},
  \bibinfo{person}{Natalie Rusk}, \bibinfo{person}{Evelyn Eastmond},
  \bibinfo{person}{Karen Brennan}, \bibinfo{person}{Amon Millner},
  \bibinfo{person}{Eric Rosenbaum}, \bibinfo{person}{Jay Silver},
  \bibinfo{person}{Brian Silverman}, {and} \bibinfo{person}{Yasmin Kafai}.}
  \bibinfo{year}{2009}\natexlab{}.
\newblock \showarticletitle{Scratch: Programming for All}.
\newblock \bibinfo{journal}{\emph{Communications of the ACM (CACM)}}
  \bibinfo{volume}{52}, \bibinfo{number}{11} (\bibinfo{date}{Nov.}
  \bibinfo{year}{2009}), \bibinfo{pages}{60--67}.
\newblock
\urldef\tempurl%
\url{https://doi.org/10.1145/1592761.1592779}
\showURL{%
\tempurl}


\bibitem[\protect\citeauthoryear{Rompf and Odersky}{Rompf and Odersky}{2012}]%
        {rompf_odersky_2012}
\bibfield{author}{\bibinfo{person}{Tiark Rompf} {and} \bibinfo{person}{Martin
  Odersky}.} \bibinfo{year}{2012}\natexlab{}.
\newblock \showarticletitle{Lightweight Modular Staging: A Pragmatic Approach
  to Runtime Code Generation and Compiled {DSLs}}.
\newblock \bibinfo{journal}{\emph{Communications of the ACM (CACM)}}
  \bibinfo{volume}{55} (\bibinfo{year}{2012}), \bibinfo{pages}{121--130}.
\newblock
Issue 6.
\urldef\tempurl%
\url{https://doi.org/10.1145/2184319.2184345}
\showURL{%
\tempurl}


\bibitem[\protect\citeauthoryear{Sahay, Indamutsa, {Di Ruscio}, and
  Pierantonio}{Sahay et~al\mbox{.}}{2020}]%
        {sahay_et_al_2020}
\bibfield{author}{\bibinfo{person}{Apurvanand Sahay}, \bibinfo{person}{Arsene
  Indamutsa}, \bibinfo{person}{Davide {Di Ruscio}}, {and}
  \bibinfo{person}{Alfonso Pierantonio}.} \bibinfo{year}{2020}\natexlab{}.
\newblock \showarticletitle{Supporting the understanding and comparison of
  low-code development platforms}. In \bibinfo{booktitle}{\emph{Euromicro
  Conference on Software Engineering and Advanced Applications (SEAA)}}.
  \bibinfo{pages}{171--178}.
\newblock
\urldef\tempurl%
\url{https://doi.org/10.1109/SEAA51224.2020.00036}
\showURL{%
\tempurl}


\bibitem[\protect\citeauthoryear{Sereshkeh, Leung, Perumal, Phillips, Zhang,
  Fazly, and Mohomed}{Sereshkeh et~al\mbox{.}}{2020}]%
        {sereshkeh_et_al_2020}
\bibfield{author}{\bibinfo{person}{Alborz~Rezazadeh Sereshkeh},
  \bibinfo{person}{Gary Leung}, \bibinfo{person}{Krish Perumal},
  \bibinfo{person}{Caleb Phillips}, \bibinfo{person}{Minfan Zhang},
  \bibinfo{person}{Afsaneh Fazly}, {and} \bibinfo{person}{Iqbal Mohomed}.}
  \bibinfo{year}{2020}\natexlab{}.
\newblock \showarticletitle{{VASTA}: A Vision and Language-Assisted Smartphone
  Task Automation System}. In \bibinfo{booktitle}{\emph{Conference on
  Intelligent User Interfaces (IUI)}}. \bibinfo{pages}{22--32}.
\newblock
\urldef\tempurl%
\url{https://doi.org/10.1145/3377325.3377515}
\showURL{%
\tempurl}


\bibitem[\protect\citeauthoryear{Shin, Lin, Thomson, Chen, Roy, Platanios,
  Pauls, Klein, Eisner, and Van~Durme}{Shin et~al\mbox{.}}{2021}]%
        {shin_et_al_2021}
\bibfield{author}{\bibinfo{person}{Richard Shin},
  \bibinfo{person}{Christopher~H. Lin}, \bibinfo{person}{Sam Thomson},
  \bibinfo{person}{Charles Chen}, \bibinfo{person}{Subhro Roy},
  \bibinfo{person}{Emmanouil~Antonios Platanios}, \bibinfo{person}{Adam Pauls},
  \bibinfo{person}{Dan Klein}, \bibinfo{person}{Jason Eisner}, {and}
  \bibinfo{person}{Benjamin Van~Durme}.} \bibinfo{year}{2021}\natexlab{}.
\newblock \showarticletitle{Constrained Language Models Yield Few-Shot Semantic
  Parsers}. In \bibinfo{booktitle}{\emph{Conference on Empirical Methods in
  Natural Language Processing (EMNLP)}}. \bibinfo{pages}{7699--7715}.
\newblock
\urldef\tempurl%
\url{https://doi.org/10.18653/v1/2021.emnlp-main.608}
\showURL{%
\tempurl}


\bibitem[\protect\citeauthoryear{{van der Aalst}, Bichler, and Heinzl}{{van der
  Aalst} et~al\mbox{.}}{2018}]%
        {vanderaalst_bichler_heinzl_2018}
\bibfield{author}{\bibinfo{person}{Wil M.~P. {van der Aalst}},
  \bibinfo{person}{Martin Bichler}, {and} \bibinfo{person}{Armin Heinzl}.}
  \bibinfo{year}{2018}\natexlab{}.
\newblock \showarticletitle{Robotic Process Automation}.
\newblock \bibinfo{journal}{\emph{Business \& Information Systems Engineering
  (BISE)}}  \bibinfo{volume}{60} (\bibinfo{year}{2018}),
  \bibinfo{pages}{269--272}.
\newblock
\urldef\tempurl%
\url{https://doi.org/10.1007/s12599-018-0542-4}
\showURL{%
\tempurl}


\bibitem[\protect\citeauthoryear{Vaziri, Mandel, Shinnar, Sim{\'e}on, and
  Hirzel}{Vaziri et~al\mbox{.}}{2017}]%
        {vaziri_et_al_2017}
\bibfield{author}{\bibinfo{person}{Mandana Vaziri}, \bibinfo{person}{Louis
  Mandel}, \bibinfo{person}{Avraham Shinnar}, \bibinfo{person}{J{\'e}r\^{o}me
  Sim{\'e}on}, {and} \bibinfo{person}{Martin Hirzel}.}
  \bibinfo{year}{2017}\natexlab{}.
\newblock \showarticletitle{Generating Chat Bots from Web API Specifications}.
  In \bibinfo{booktitle}{\emph{Symposium on New Ideas, New Paradigms, and
  Reflections on Programming and Software (Onward!)}}. \bibinfo{pages}{44--57}.
\newblock
\urldef\tempurl%
\url{http://doi.acm.org/10.1145/3133850.3133864}
\showURL{%
\tempurl}


\bibitem[\protect\citeauthoryear{Voelter and Lisson}{Voelter and
  Lisson}{2014}]%
        {voelter_lisson_2014}
\bibfield{author}{\bibinfo{person}{Markus Voelter} {and}
  \bibinfo{person}{Sascha Lisson}.} \bibinfo{year}{2014}\natexlab{}.
\newblock \showarticletitle{Supporting Diverse Notations in {MPS}' Projectional
  Editor.}. In \bibinfo{booktitle}{\emph{Workshop on The Globalization of
  Modeling Languages (GEMOC)}}. \bibinfo{pages}{7--16}.
\newblock
\urldef\tempurl%
\url{https://hal.inria.fr/hal-01074602/file/GEMOC2014-complete.pdf#page=13}
\showURL{%
\tempurl}


\bibitem[\protect\citeauthoryear{Wang, Berant, and Liang}{Wang
  et~al\mbox{.}}{2015}]%
        {wang_berant_liang_2015}
\bibfield{author}{\bibinfo{person}{Yushi Wang}, \bibinfo{person}{Jonathan
  Berant}, {and} \bibinfo{person}{Percy Liang}.}
  \bibinfo{year}{2015}\natexlab{}.
\newblock \showarticletitle{Building a semantic parser overnight}. In
  \bibinfo{booktitle}{\emph{Annual Meeting of the Association for Computational
  Linguistics (ACL)}}. \bibinfo{pages}{1332--1342}.
\newblock
\urldef\tempurl%
\url{https://www.aclweb.org/anthology/P15-1129.pdf}
\showURL{%
\tempurl}


\bibitem[\protect\citeauthoryear{Yin and Neubig}{Yin and Neubig}{2017}]%
        {yin_neubig_2017}
\bibfield{author}{\bibinfo{person}{Pengcheng Yin} {and} \bibinfo{person}{Graham
  Neubig}.} \bibinfo{year}{2017}\natexlab{}.
\newblock \showarticletitle{A Syntactic Neural Model for General-Purpose Code
  Generation}. In \bibinfo{booktitle}{\emph{Annual Meeting of the Association
  for Computational Linguistics (ACL)}}. \bibinfo{pages}{440--450}.
\newblock
\urldef\tempurl%
\url{http://dx.doi.org/10.18653/v1/P17-1041}
\showURL{%
\tempurl}


\bibitem[\protect\citeauthoryear{Zhang, Zhang, and Cao}{Zhang
  et~al\mbox{.}}{2001}]%
        {zhang_zhang_cao_2001}
\bibfield{author}{\bibinfo{person}{K. Zhang}, \bibinfo{person}{D.-Q. Zhang},
  {and} \bibinfo{person}{J. Cao}.} \bibinfo{year}{2001}\natexlab{}.
\newblock \showarticletitle{Design, construction, and application of a generic
  visual language generation environment}.
\newblock \bibinfo{journal}{\emph{IEEE Transactions on Software Engineering
  (TSE)}} \bibinfo{volume}{27}, \bibinfo{number}{4} (\bibinfo{year}{2001}),
  \bibinfo{pages}{289--307}.
\newblock
\urldef\tempurl%
\url{https://doi.org/10.1109/32.917521}
\showURL{%
\tempurl}


\end{thebibliography}

\end{document}